\documentclass{blois}

\usepackage{amsmath}
\usepackage{graphicx}
\usepackage{caption}

\def\be{\begin{equation}}
\def\ee{\end{equation}}
\def\bea{\begin{eqnarray}}
\def\eea{\end{eqnarray}}

\newcommand{\Photo}{}

\begin{document}
\hfill HU-EP-19/2
\vspace*{4cm}
\title{Cross section ratios as a precision tool for $t\bar{t}\gamma$}

\author{ Manfred Kraus }

\address{Humboldt-Universit\"at zu Berlin, Institut f\"ur Physik\\ Newtonstra\ss{}e 15, D-12489 Berlin, Germany}

\maketitle\abstracts{
We report on our recent calculation for the off-shell $t\bar{t}\gamma$ process and its 
potential for precision measurements by constructing ratios of total and differential
cross sections. Precise theoretical predictions for these ratios can help to constrain
new physics contributions in the top-quark sector of the Standard Model.}

\section{Introduction}
\begin{figure}[h!]
\begin{center}
 \includegraphics[width=0.55\textwidth]{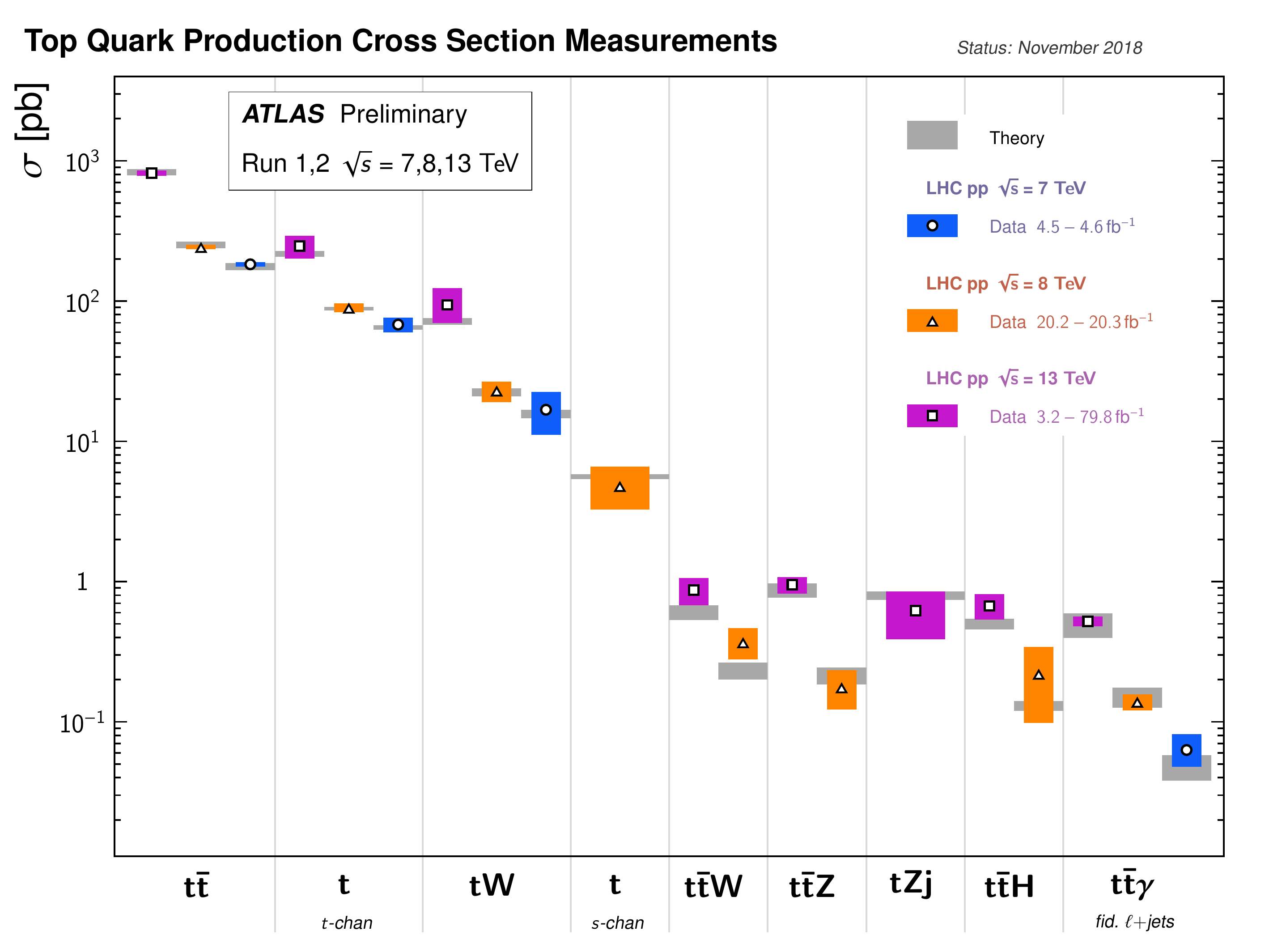}
 \caption{Cross sections for various top-quark processes. Taken from Ref.~\protect\cite{ATL-PHYS-PUB-2018-034}}.
 \label{fig:ATLAS_xsecs}
\end{center}
\end{figure}

Top quarks are abundantly produced in pairs via strong interactions at the 
Large Hadron Collider (LHC) and are subject of precision measurements.
Besides the dominant top-quark pair production process the LHC, running at a 
center-of-mass energy of $\sqrt{s} = 13$ TeV, offers the possibility to 
investigate also rare top-quark processes such as single top-quark production
or associated top-quark pair production with additional bosons. 
In Fig.~\ref{fig:ATLAS_xsecs} the cross section measurements of various production
channels are summarized.
In particular the detailed study of top-quark pair production in association with 
a hard photon opens up a new window for precision measurements of top-quark 
properties such as the top-quark electric charge, its electroweak dipole moments
\cite{Schulze:2016qas} or the top-quark charge asymmetry~\cite{Bergner:2018lgm}.

Due to the rich phenomenology of the $t\bar{t}\gamma$ final state this process 
has received a lot of attention in recent years. For on-shell top-quarks the NLO 
QCD corrections were computed first in Refs.~\cite{PengFei:2009ph,PengFei:2011qg,Maltoni:2015ena},
while electroweak corrections became available~\cite{Duan:2016qlc} only recently.
For on-shell tops also the matching to parton showers has been achieved via 
the POWHEG method~\cite{Kardos:2014zba}. At fixed-order in NLO also the corrections 
to the top-quark decay have been incorporated via the Narrow-width-approximation~\cite{Melnikov:2011ta}
and only recently the full off-shell calculation in the dileptonic decay channel has 
been presented in Refs.~\cite{Bevilacqua:2018woc,Bevilacqua:2018dny}.
On the experimental side also strong efforts are being made to isolate the $t\bar{t}\gamma$
process. Already in 2011, the first evidence had been reported by the CDF Collaboration~\cite{Aaltonen:2011sp} at the TeVatron.
Afterwards, studies were continued at the LHC and the first observation at $\sqrt{s}=7$ TeV has been 
made~\cite{Aad:2015uwa} with subsequent analyses performed at $\sqrt{s} = 8$ TeV~\cite{Aaboud:2017era,Sirunyan:2017iyh} 
and $\sqrt{s} = 13$ TeV~\cite{Aaboud:2018hip}.
 
In this proceeding we are discussing the recent results~\cite{Bevilacqua:2018dny} for precise cross section 
ratios for top-quark pair production in association with a photon in the fiducial 
phase space volume
\begin{equation}
  \mathcal{R} = \frac{\sigma_{pp \to e^+\nu_e \mu^-\bar{\nu}_\mu b\bar{b}\gamma}}{\sigma_{pp \to e^+\nu_e \mu^-\bar{\nu}_\mu b\bar{b}}}\;.
\end{equation}
From the experimental point of view these cross section ratios offer a possibility 
to reduce the impact of systematic uncertainties since they can cancel between the 
numerator and denominator. However, for theoretical predictions these cancellations
are not guaranteed. For example, the dependence on the unphysical renormalisation and 
factorisation scales is expected to decrease only if the processes involved in the ratio 
are highly correlated. 
\begin{figure}[h!]
\begin{center}
 \includegraphics[width=0.49\textwidth]{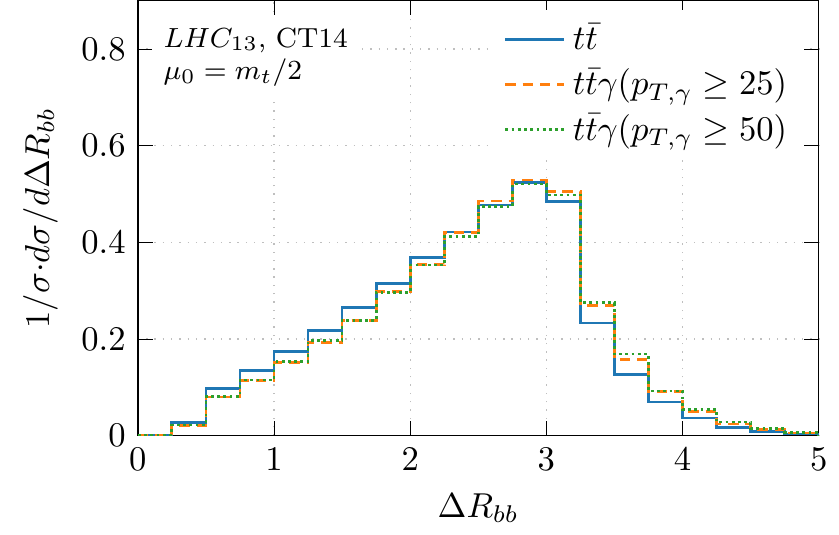}
 \includegraphics[width=0.49\textwidth]{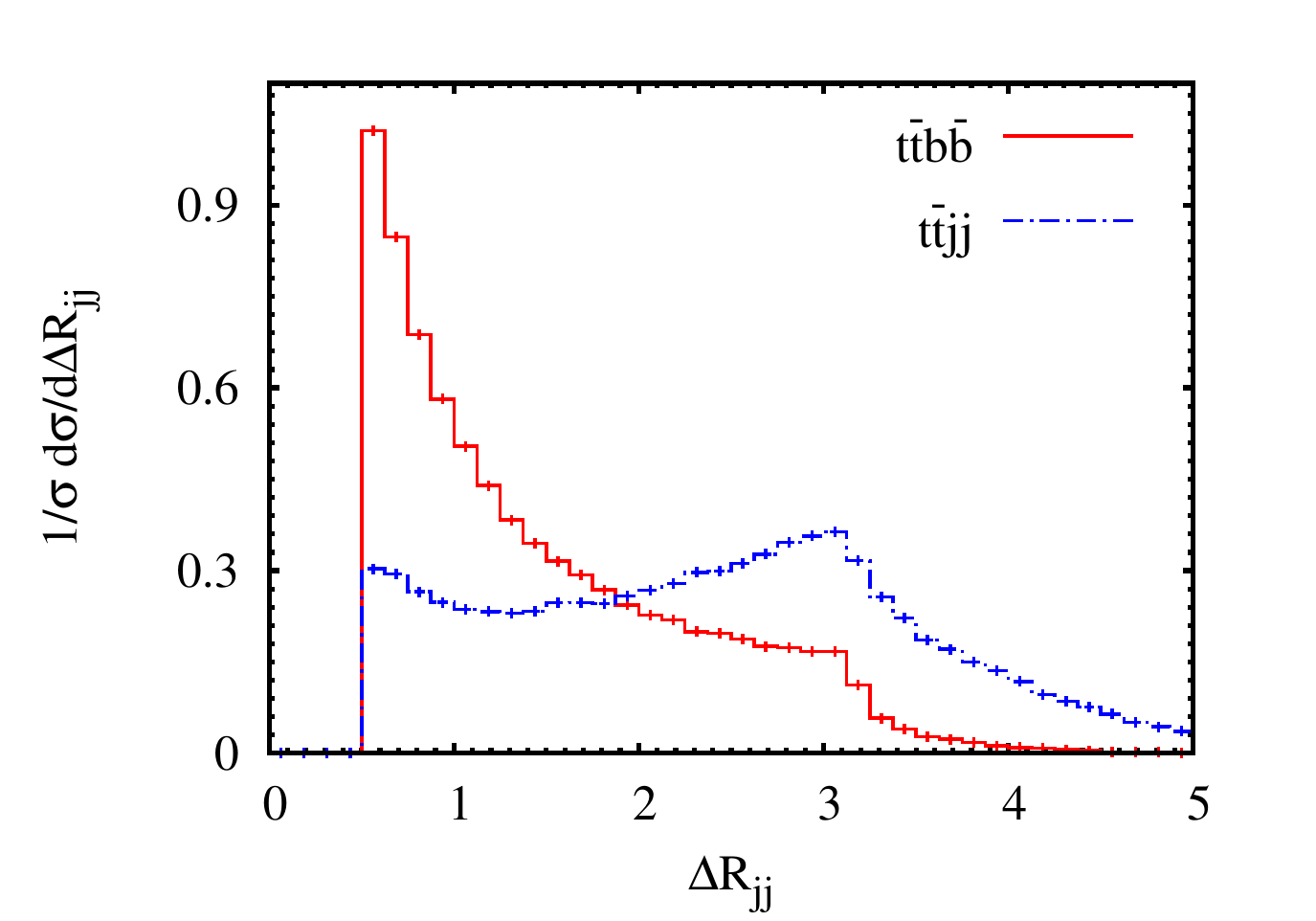}
 \caption{Shape differences for differential cross sections. Left: $t\bar{t}\gamma$ vs. $t\bar{t}$ Right: $t\bar{t}jj$ vs. $t\bar{t}b\bar{b}$}.
 \label{fig:correlations}
\end{center}
\end{figure}
In Fig.~\ref{fig:correlations} we show an example for a correlated and an uncorrelated process.
On the left plot the shape comparison of $t\bar{t}$ and $t\bar{t}\gamma$ is shown for the 
observable $\Delta R_{bb}$. We observe that the shape agrees well even for high transverse momentum 
cuts on the additional hard photon. Therefore, we can conclude that $t\bar{t}$ and $t\bar{t}\gamma$
are indeed correlated and should receive similar QCD corrections. A more thorough comparison can be found in Ref.~\cite{Bevilacqua:2018dny}.
On the other hand, we present also 
a comparison from Ref.~\cite{Bevilacqua:2014qfa} where the correlations between $t\bar{t}b\bar{b}$ and $t\bar{t}jj$
were investigated. Here we observe large differences in the differential distributions and therefore
these processes can not be treated as correlated and a cancellation of the scale dependence can not be 
expected.
\section{Cross section ratios}
The calculation is performed using the \texttt{HELAC-NLO} framework~\cite{Bevilacqua:2011xh}
that consists out of \texttt{HELAC-1Loop}~\cite{vanHameren:2009dr} and 
\texttt{HELAC-Dipoles}~\cite{Czakon:2009ss,Bevilacqua:2013iha}. The matrix elements
are based on the complete final state $e^+\nu_e \mu^-\bar{\nu}_\mu b\bar{b}\gamma$
including all resonant and non-resonant contributions as well as off-shell and 
interference effects. Results are stored as ROOT Ntuple event files, where additional
information is kept, such that a reweighting to different scales or PDFs can easily
be accomplished~\cite{Bern:2013zja}. The framework has been previously used to 
obtain results for various off-shell top-quark processes~\cite{Bevilacqua:2010qb,Bevilacqua:2015qha,Bevilacqua:2016jfk,Bevilacqua:2017ipv,Bevilacqua:2019cvp}.
Further details on input parameters and selection cuts can be found in Refs.~\cite{Bevilacqua:2018woc,Bevilacqua:2018dny}.

We study cross section ratios for $pp \to e^+\nu_e \mu^-\bar{\nu}_\mu b\bar{b}\gamma$
with respect to $pp \to e^+\nu_e \mu^-\bar{\nu}_\mu b\bar{b}$ at the LHC at a 
center-of-mass energy of $\sqrt{s} = 13$ TeV. The ratio of the fiducial cross sections $\mathcal{R}$ 
are found to be
\begin{equation}
\begin{split}	
  \mathcal{R}(\mu_0 = H_T/4, p_{T,\gamma} > 25~\textrm{GeV}) &= (4.62 \pm 0.06~\textrm{[scales]} \pm 0.02~\textrm{[PDFs]})\cdot 10^{-3}\;, \\
  \mathcal{R}(\mu_0 = H_T/4, p_{T,\gamma} > 50~\textrm{GeV}) &= (1.93 \pm 0.06~\textrm{[scales]} \pm 0.02~\textrm{[PDFs]})\cdot 10^{-3}\,.
\end{split}	
\end{equation}
The resulting theoretical uncertainty stemming from the scale variation amounts to roughly $2-3\%$, which is smaller than 
the individual cross section uncertainties. Thus a partial cancellation of the scale variation can be observed.
However, the uncertainties related to the scale choice still dominate over the PDF uncertainties.
The dramatic reduction of the scale variation can also be observed in differential cross section ratios.
\begin{figure}[h!]
 \includegraphics[width=0.49\textwidth]{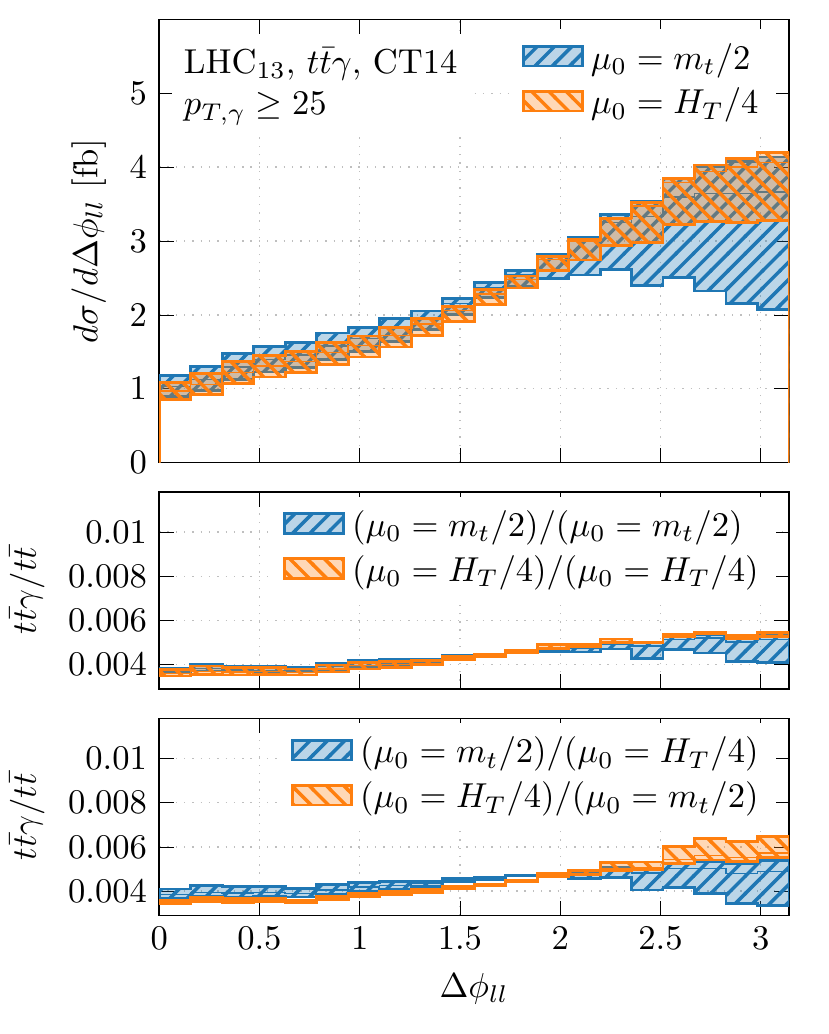}
 \includegraphics[width=0.49\textwidth]{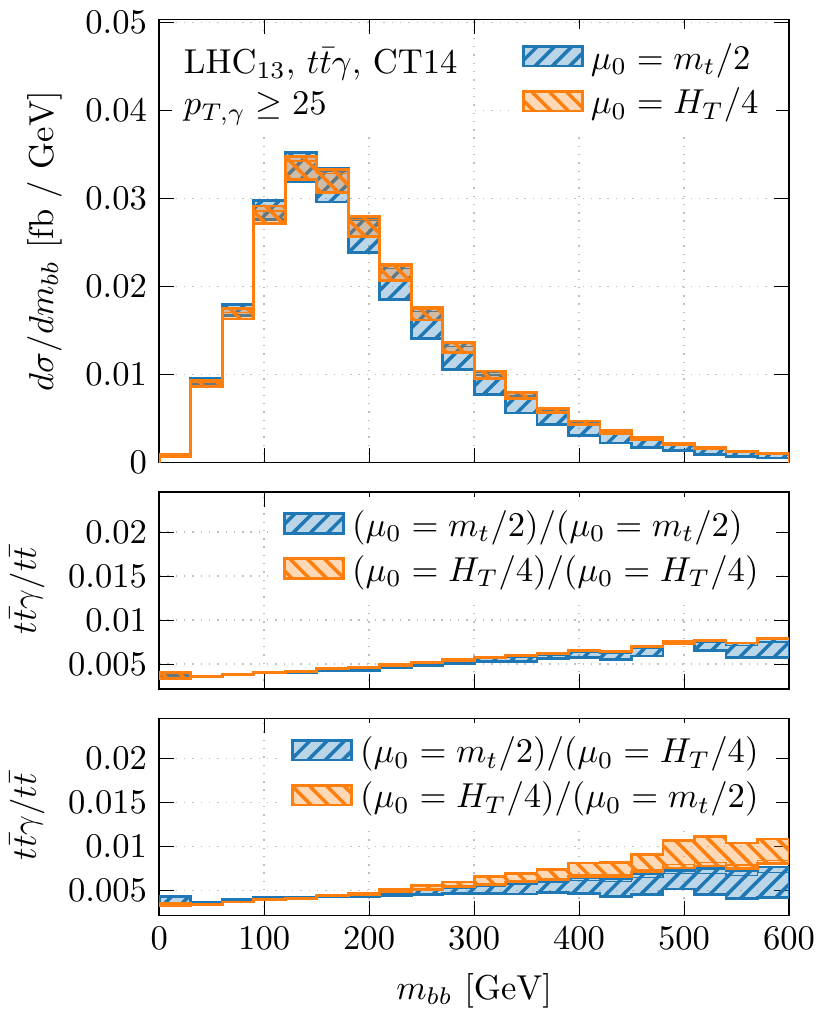}
 \caption{Differential cross section ratios for $\Delta\phi_{\ell\ell}$ and the invariant mass of the b-jets $M_{bb}$.}
 \label{fig:diffXS}
\end{figure}
Fig.~\ref{fig:diffXS} depicts the azimuthal angle between the leptons and the invariant mass of the two b-jets. The upper panel 
shows the absolute predictions for $t\bar{t}\gamma$, while the lower panels show the correlated and uncorrelated ratios. For 
both observables we notice that the dynamical scale yields a smaller scale variation for the absolute predictions as compared to the 
fixed scale choice. For example,
in the case of $\Delta\phi_{\ell\ell}$ a reduction of the uncertainty by more than a factor of $2$ can be observed at the end of the
spectrum. Considering the differential ratios in the bottom panels we see that the correlated ratio, i.e. the same scale is used in the numerator and 
denominator, the ratio can be stabilised with a residual uncertainty of the order of $3\%$. On the other hand, if an uncorrelated 
scale choice has been made, the ratio yields a much larger uncertainties as before.
Even for hadronic observables such as the invariant mass of the b-jets, the ratio becomes very precise, as can be seen on the right plot of 
Fig.~\ref{fig:diffXS}. Also here the same scale choice for the $t\bar{t}$ and $t\bar{t}\gamma$ yields the smallest residual scale uncertainties.

\section{Conclusions}
We have investigated the potential precision of cross section ratios for $t\bar{t}\gamma$ and $t\bar{t}$ in the dileptonic decay
channel including NLO QCD corrections. Due to the similarity of the processes a correlated scale choice reduces significantly the 
residual theoretical uncertainties. However, uncertainties stemming from scale variations are still the dominant source. The
obtained (differential) cross section ratios can be used to constrain new physics contributions or to probe the top-quark interaction
with the photon with high precision.

\section*{References}

\end{document}